# $\mathbf{N=4}$ Supersymmetric Yang-Mills Multiplet in Non-Adjoint Representations

Hitoshi NISHINO[1] and Subhash RAJPOOT[2]

*Department of Physics & Astronomy*
*California State University*
*1250 Bellflower Boulevard*
*Long Beach, CA 90840*

**Abstract**

We formulate a theory for $N = 4$ supersymmetric Yang-Mills multiplet in a non-adjoint representation $R$ of $SO(\mathcal{N})$, as an important application of our recently-proposed model for $N = 1$ supersymmetry. This system is obtained by dimensional reduction from an $N = 1$ supersymmetric Yang-Mills multiplet in non-adjoint representation in ten dimensions. The consistency with supersymmetry requires that the non-adjoint representation $R$ with the indices ${\scriptstyle i,~j,~\cdots}$ satisfy the three conditions $\eta^{ij} = \delta^{ij}$, $(T^I)^{ij} = -(T^I)^{ji}$ and $(T^I)^{[ij|}(T^I)^{|k]l} = 0$ for the metric $\eta^{ij}$ and the generators $T^I$, which are the same as the $N = 1$ case.

[1] E-Mail: hnishino@csulb.edu

[2] E-Mail: rajpoot@csulb.edu

## 1. Introduction

The importance of $N = 4$ extended supersymmetry in four-dimensions (4D) [1] is associated with its all-order finiteness [2], and also its natural link with superstring theories in 4D [3]. Moreover, there is an important duality between $N = 4$ supersymmetric Yang-Mills theory in 4D and IIB string theory in 10D compactified on $\mathrm{AdS}_5 \times \mathrm{S}_5$ [4][3)]

In the conventional formulation of $N = 1$ supersymmetry in 4D, a vector multiplet is supposed to be in the adjoint representation, such as $(A_\mu{}^I, \lambda^I)$ carrying the common adjoint index $I$ [1]. However, we have recently shown [6] that this is not necessarily the case, by constructing an explicit $N = 1$ Yang-Mills multiplet in a non-adjoint representation. We have shown that the multiplet $(B_\mu{}^i, \chi^i)$ with the non-adjoint real representation index $i$ can consistently couple to the conventional Yang-Mills multiplet $(A_\mu{}^I, \lambda^I)$. Such a non-adjoint real representation $R$ should satisfy certain conditions [6] for the system to be consistent with supersymmetry (Cf. (2.1) below).

In this paper, we show that the $N = 1$ formulation in [6] can be further generalized to extended $N = 4$ supersymmetry. In addition to the conventional $N = 4$ supersymmetric Yang-Mills multiplet $(A_\mu{}^I, \lambda_{(i)}{}^I, A_\alpha{}^I, \widetilde{A}_\alpha{}^I)$ $(\alpha = 1,\, 2,\, 3;\ \ (i) = 1,\, 2,\, 3,\, 4)$, we can consider the additional vector multiplet $(B_\mu{}^i, \chi_{(i)}{}^i, B_\alpha{}^i, \widetilde{B}_\alpha{}^i)$ carrying the index $i$ for a particular non-adjoint real representation $R$. As explained in the case of $N = 1$ [6], we have to maintain the conventional Yang-Mills in the adjoint representation, once we introduce the extra vector multiplet in the non-adjoint representation $R$.

It seems to be a prevailing notion that $N = 4$ supersymmetric Yang-Mills theory in 4D has the 'unique' field content all in the adjoint representation of a certain gauge group. For example, the first sentence of section 3 in [5] states that "The Lagrangian for the $N = 4$ super-Yang Mills theory is unique". In our present paper, we establish a counter-example against the prevailing notion of the 'uniqueness' of $N = 4$ supersymmetric Yang-Mills theory in 4D.

## 2. The Lagrangian

As has been mentioned, our system has two $N = 4$ vector multiplets $(A_\mu{}^I, \lambda_{(i)}{}^I, A_\alpha{}^I, \widetilde{A}_\alpha{}^I)$ and $(B_\mu{}^i, \chi_{(i)}{}^i, B_\alpha{}^i, \widetilde{B}_\alpha{}^i)$. The indices $(i),\, (j),\, \cdots = 1,\, 2,\, 3,\, 4$ are for $N = 4$ supersymmetry, while the indices $\alpha,\, \beta,\, \cdots = 1,\, 2,\, 3$ are used for the three scalars and three pseudo-scalars [1].

3) For reviews, see, e.g., ref. [5].

The former multiplet is the conventional $N = 4$ supersymmetric Yang-Mills vector multiplet [1] with the adjoint index $I$ of the gauge group $SO(\mathcal{N})$. The latter multiplet is our new vector multiplet carrying the indices $i, j, \cdots$ for the non-adjoint representation $R$ of $SO(\mathcal{N})$, which satisfies the conditions

$$\eta^{ij} = \delta^{ij} \quad , \qquad (T^I)^{ij} = -(T^I)^{ji} \quad , \tag{2.1a}$$

$$(T^I)^{[ij|}(T^I)^{|k]l} \equiv 0 \quad , \tag{2.1b}$$

where $\eta^{ij}$ and $(T^I)^{ij}$ are the metric of the representation $R$, and the representation matrix of the generators of $SO(\mathcal{N})$, respectively.

We can obtain our lagrangian for $N = 4$ supersymmetry for these two multiplets by the simple dimensional reduction [7] of $N = 1$ supersymmetric Yang-Mills in 10D in the non-adjoint representation, as outlined in [6]. Our lagrangian thus obtained in 4D is

$$\begin{aligned}
\mathcal{L} = & -\tfrac{1}{4}(\mathcal{F}_{\mu\nu}{}^I)^2 - \tfrac{1}{4}(G_{\mu\nu}{}^i)^2 + \tfrac{1}{2}(\overline{\lambda}{}^I \not{\mathcal{D}} \lambda^I) + \tfrac{1}{2}(\overline{\chi}{}^i \not{\mathcal{D}} \chi^i) \\
& - \tfrac{1}{2}(\mathcal{D}_\mu A_\alpha{}^I)^2 - \tfrac{1}{2}(\mathcal{D}_\mu \widetilde{A}_\alpha{}^I)^2 - \tfrac{1}{2}(\mathcal{D}_\mu B_\alpha{}^i)^2 - \tfrac{1}{2}(\mathcal{D}_\mu \widetilde{B}_\alpha{}^i)^2 \\
& - \tfrac{i}{2} g f^{IJK} (\overline{\lambda}{}^I \alpha_\alpha \lambda^J) A_\alpha{}^K - \tfrac{1}{2} g f^{IJK} (\overline{\lambda}{}^I \gamma_5 \beta_\alpha \lambda^J) \widetilde{A}_\alpha{}^K \\
& + i g (T^I)^{ij} (\overline{\lambda}{}^I \alpha_\alpha \chi^i) B_\alpha{}^j + g (T^I)^{ij} (\overline{\lambda}{}^I \gamma_5 \beta_\alpha \chi^i) \widetilde{B}_\alpha{}^j \\
& + \tfrac{i}{2} (T^I)^{ij} (\overline{\chi}{}^i \alpha_\alpha \chi^j) A_\alpha{}^I + \tfrac{1}{2} (T^I)^{ij} (\overline{\chi}{}^i \gamma_5 \beta_\alpha \chi^j) \widetilde{A}_\alpha{}^I \\
& - \tfrac{1}{4} g^2 \left[ f^{IJK} A_\alpha{}^J A_\beta{}^K - (T^I)^{ij} B_\alpha{}^i B_\beta{}^j \right]^2 - \tfrac{1}{4} g^2 \left[ f^{IJK} \widetilde{A}_\alpha{}^J \widetilde{A}_\beta{}^K - (T^I)^{ij} \widetilde{B}_\alpha{}^i \widetilde{B}_\beta{}^j \right]^2 \\
& - \tfrac{1}{2} g^2 \left[ f^{IJK} A_\alpha{}^J \widetilde{A}_\beta{}^K - (T^I)^{ij} B_\alpha{}^i \widetilde{B}_\beta{}^j \right]^2 \\
& - \tfrac{1}{4} g^2 \left[ (T^I)^{ij} (A_\alpha{}^I B_\beta{}^j - A_\beta{}^I B_\alpha{}^j) \right]^2 - \tfrac{1}{4} g^2 \left[ (T^I)^{ij} (\widetilde{A}_\alpha{}^I \widetilde{B}_\beta{}^j - \widetilde{A}_\beta{}^I \widetilde{B}_\alpha{}^j) \right]^2 \\
& - \tfrac{1}{2} g^2 \left[ (T^I)^{ij} (A_\alpha{}^I \widetilde{B}_\beta{}^j - \widetilde{A}_\beta{}^I B_\alpha{}^j) \right]^2 \quad .
\end{aligned} \tag{2.2}$$

The $4 \times 4$ antisymmetrtic matrices $\alpha$ and $\beta$ satisfy the conditions for $\alpha, \beta, \cdots = 1, 2, 3$:

$$\alpha_\alpha \alpha_\beta = \delta_{\alpha\beta} + i\epsilon_{\alpha\beta\gamma}\alpha_\gamma \quad , \qquad \beta_\alpha \beta_\beta = \delta_{\alpha\beta} + i\epsilon_{\alpha\beta\gamma}\beta_\gamma \quad , \qquad [\alpha_\alpha, \beta_\beta] = 0 \quad , \tag{2.3}$$

which are $SO(3)$ matrices, and used for the global $SO(4) \approx SO(3) \times SO(3)$ [1]. Similarly to [6], our field strengths and covariant derivatives are defined by

$$\mathcal{F}_{\mu\nu}{}^I \equiv 2\partial_{[\mu} A_{\nu]}{}^I + g f^{IJK} A_\mu{}^J A_\nu{}^K - g (T^I)^{ij} B_\mu{}^i B_\nu{}^j \quad , \tag{2.4a}$$

$$G_{\mu\nu}{}^i \equiv 2\partial_{[\mu} B_{\nu]}{}^i + 2g (T^I)^{ij} A_{[\mu}{}^I B_{\nu]}{}^j \quad , \tag{2.4b}$$

$$\mathcal{D}_\mu \chi^i \equiv \partial_\mu \chi^i + g (T^I)^{ij} A_\mu{}^I \chi^j - g (T^I)^{ij} B_\mu{}^j \lambda^I \quad , \tag{2.4c}$$

$$\mathcal{D}_\mu \lambda^I \equiv \partial_\mu \lambda^I + g f^{IJK} A_\mu{}^J \lambda^K - g(T^I)^{ij} B_\mu{}^i \chi^j \quad , \tag{2.4d}$$

$$\mathcal{D}_\mu A_\alpha{}^I \equiv \partial_\mu A_\alpha{}^I + g f^{IJK} A_\mu{}^J A_\alpha{}^K - g(T^I)^{ij} B_\mu{}^i B_\alpha{}^j \quad , \tag{2.4e}$$

$$\mathcal{D}_\mu \widetilde{A}_\alpha{}^I \equiv \partial_\mu \widetilde{A}_\alpha{}^I + g f^{IJK} A_\mu{}^J \widetilde{A}_\alpha{}^K - g(T^I)^{ij} B_\mu{}^i \widetilde{B}_\alpha{}^j \quad , \tag{2.4f}$$

$$\mathcal{D}_\mu B_\alpha{}^i \equiv \partial_\mu B_\alpha{}^i + g(T^I)^{ij} A_\mu{}^I B_\alpha{}^j - g(T^I)^{ij} A_\alpha{}^I B_\mu{}^j \quad , \tag{2.4g}$$

$$\mathcal{D}_\mu \widetilde{B}_\alpha{}^i \equiv \partial_\mu \widetilde{B}_\alpha{}^i + g(T^I)^{ij} A_\mu{}^I \widetilde{B}_\alpha{}^j - g(T^I)^{ij} \widetilde{A}_\alpha{}^I B_\mu{}^j \quad . \tag{2.4h}$$

Our action $I \equiv \int d^4x \, \mathcal{L}$ is invariant under supersymmetry

$$\delta_Q A_\mu{}^I = + (\overline{\epsilon}\gamma_\mu \lambda^I) \quad , \qquad \delta_Q B_\mu{}^i = +(\overline{\epsilon}\gamma_\mu \chi^i) \quad , \tag{2.5a}$$

$$\delta_Q A_\alpha{}^I = + i(\overline{\epsilon}\alpha_\alpha \lambda^I) \quad , \qquad \delta_Q B_\alpha{}^I = +i(\overline{\epsilon}\alpha_\alpha \chi^i) \quad , \tag{2.5b}$$

$$\delta_Q \widetilde{A}_\alpha{}^I = + (\overline{\epsilon}\gamma_5 \beta_\alpha \lambda^I) \quad , \qquad \delta_Q \widetilde{B}_\alpha{}^I = +(\overline{\epsilon}\gamma_5 \beta_\alpha \chi^i) \quad , \tag{2.5c}$$

$$\begin{aligned} \delta_Q \lambda^I = &+ \tfrac{1}{2}(\gamma^{\mu\nu}\epsilon)\mathcal{F}_{\mu\nu}{}^I + i(\alpha_\alpha \gamma^\mu \epsilon)\mathcal{D}_\mu A_\alpha{}^I - (\beta_\alpha\gamma_5\gamma^\mu\epsilon)\mathcal{D}_\mu \widetilde{A}_\alpha{}^I \\ &+ \tfrac{i}{2} g \epsilon_{\alpha\beta\gamma}(\alpha_\gamma \epsilon) \left[ f^{IJK} A_\alpha{}^J A_\beta{}^K - (T^I)^{ij} B_\alpha{}^i B_\beta{}^j \right] \\ &+ \tfrac{i}{2} g \epsilon_{\alpha\beta\gamma}(\beta_\gamma \epsilon) \left[ f^{IJK} \widetilde{A}_\alpha{}^J \widetilde{A}_\beta{}^K - (T^I)^{ij} \widetilde{B}_\alpha{}^i \widetilde{B}_\beta{}^j \right] \\ &- ig(\alpha_\alpha \beta_\beta \gamma_5 \epsilon) \left[ f^{IJK} A_\alpha{}^J \widetilde{A}_\beta{}^K - (T^I)^{ij} B_\alpha{}^i \widetilde{B}_\beta{}^j \right] \quad , \end{aligned} \tag{2.5d}$$

$$\begin{aligned} \delta_Q \chi^i = &+ \tfrac{1}{2}(\gamma^{\mu\nu}\epsilon) G_{\mu\nu}{}^i + i(\alpha_\alpha \gamma^\mu \epsilon)\mathcal{D}_\mu B_\alpha{}^i - (\beta_\alpha\gamma_5\gamma^\mu\epsilon)\mathcal{D}_\mu \widetilde{B}_\alpha{}^i \\ &+ ig\epsilon_{\alpha\beta\gamma}(T^I)^{ij}(\alpha_\gamma\epsilon) A_\alpha{}^I B_\beta{}^j + ig\epsilon_{\alpha\beta\gamma}(T^I)^{ij}(\beta_\gamma\epsilon)\widetilde{A}_\alpha{}^I \widetilde{B}_\beta{}^j \\ &- ig(\alpha_\alpha\beta_\beta\gamma_5\epsilon)(T^I)^{ij}(A_\alpha{}^I \widetilde{B}_\beta{}^j - \widetilde{A}_\beta{}^I B_\alpha{}^j) \quad . \end{aligned} \tag{2.5e}$$

The supersymmetric invariance of our action $\delta_Q I = 0$ can be confirmed in the usual way. The crucial relationships are the conditions (2.1), as well as the Bianchi identities

$$\mathcal{D}_{[\mu} \mathcal{F}_{\nu\rho]}{}^I \equiv \partial_{[\mu} \mathcal{F}_{\nu\rho]}{}^I + g f^{IJK} A_{[\mu}{}^J \mathcal{F}_{\nu\rho]}{}^K - g(T^I)^{ij} B_{[\mu}{}^i G_{\nu\rho]}{}^j \equiv 0 \quad , \tag{2.6a}$$

$$\mathcal{D}_{[\mu} G_{\nu\rho]}{}^i \equiv \partial_{[\mu} G_{\nu\rho]}{}^i + g(T^I)^{ij} A_{[\mu}{}^I G_{\nu\rho]}{}^j - g(T^I)^{ij} B_{[\mu}{}^j \mathcal{F}_{\nu\rho]}{}^I \equiv 0 \quad . \tag{2.6b}$$

At the cubic-order level in $\delta_Q \mathcal{L}$, we need the Fierz identity

$$\begin{aligned} \Big[ (\gamma_\mu)_{AB}(\gamma^\mu)_{CD} \, \delta_{(i)(j)}\delta_{(k)(\ell)} - C_{AB}C_{CD}(\alpha_\alpha)_{(i)(j)}(\alpha_\alpha)_{(k)(\ell)} \\ + (\gamma_5)_{AB}(\gamma_5)_{CD}(\beta_\alpha)_{(i)(j)}(\beta_\alpha)_{(k)(\ell)} \Big] + (2 \text{ perm.}) \equiv 0 \quad , \end{aligned} \tag{2.7}$$

where $A, \, B, \, \cdots = 1, \, \cdots, \, 4$ are for the Majorana spinor components in 4D, while '2 perm.' stands for the two more sets of terms for the cyclic permutations of $B(j) \rightarrow C(k), \; C(k) \rightarrow D(\ell), \; D(\ell) \rightarrow B(j),$ so that the whole expression is totally symmetric with respect to these three pairs of indices. This identity is used both for the $g\lambda\chi^2$ and the $g\lambda^3$-terms in $\delta_Q\mathcal{L}$. The key ingredient at the quartic-order level is the usage of the condition (2.1b) in the sector $g^2\chi B_\alpha B_\beta \widetilde{B}_\gamma$ in the variation $\delta_Q\mathcal{L}$.

Our peculiar vector multiplet $(B_\mu{}^i, \lambda^i, B_\alpha{}^i, \widetilde{B}_\alpha{}^i)$ carrying the indices $i$ of the representation $R$ of $SO(\mathcal{N})$ must satisfy the conditions in (2.1). A necessary conditions of (2.1b) is [6]

$$\frac{2dI_2(R)}{N(N-1)} - 2I_2(R) + N - 2 = 0 \quad , \tag{2.8}$$

where $d \equiv \dim(R)$, while the second index $I_2(R)$ is defined by $(T^I T^I)^{ij} = -2I_2(R)\delta^{ij}$, and accordingly $(T^I T^J)^{ii} = -4dI_2(R)\delta^{IJ}/N(N-1)$. As long as these conditions are satisfied, the representation $R$ can be any real representation of $SO(\mathcal{N})$. A trivial example is the $\underset{\sim}{\mathcal{N}}$ of $SO(\mathcal{N})$, but this system has the hidden local symmetry, *i.e.,* the system is equivalent to a supersymmetric Yang-Mills theory for the local $SO(\mathcal{N}+1)$ [6]. Non-trivial examples are the $\mathbf{8_C}$ and $\mathbf{8_S}$-representations of $SO(8)$ different from the usual $\mathbf{8_V}$-representation.

## 3. Summary and Concluding Remarks

In this paper, we have constructed the system of $N=4$ supersymmetric Yang-Mills multiplet in a non-adjoint real representation $R$. As long as the conditions in (2.1) are satisfied for the representation $R$, we have found the extra vector multiplet $(B_\mu{}^i, \chi_{(i)}{}^i, B_\alpha{}^i, \widetilde{B}_\alpha{}^i)$ can be coupled to the conventional vector multiplet $(A_\mu{}^I, \lambda_{(i)}{}^I, A_\alpha{}^I, \widetilde{A}_\alpha{}^I)$ consistently with supersymmetry. As in the $N=1$ case, we need at least the conventional $N=4$ supersymmetric Yang-Mills multiplet in the adjoint representation, once we introduce the vector multiplet in the non-adjoint representation $R$. The non-trivial examples of such representations are the $\mathbf{8_S}$ and $\mathbf{8_C}$ of $SO(8)$.

According to the prevailing notion, since the $N=4$ supersymmetry is the maximal extended global supersymmetry in 4D, there is *no* other outside multiplet that can be coupled to the basic $8+8$ multiplet $(A_\mu{}^I, \lambda_{(i)}{}^I, A_\alpha{}^I, \widetilde{A}_\alpha{}^I)$. In that sense, the field content of $N=4$ supersymmetric theory is supposed to be unique [5]. However, we already know one counter-example against this wisdom, namely a vector multiplet gauging scale symmetry presented in [8]. Our theory in this paper has established another counter-example now with the $N=4$ vector multiplet in the non-adjoint representation $R$ consistently coupled to the conventional $N=4$ Yang-Mills multiplet. It is amazing that such a tight $N=4$ maximally extended supermultiplet with $8+8$ degrees of freedom can be further coupled to an extra vector multiplet with additional physical degrees of freedom.

As the conventional $N=4$ supersymmetric Yang-Mills theory is finite [2], so may well be our theory to all orders.

## References

[1] L. Brink, J.H. Schwarz and J. Scherk, Nucl. Phys. **B121** (1977) 77; F. Gliozzi, J. Scherk and D. Olive, Nucl. Phys. **B122** (1977) 253.

[2] S. Mandelstam, Nucl. Phys. **B213** (1983) 149; L. Brink, O. Lindgren and B.E.W. Nilsson, Phys. Lett. **123B** (1983) 323.

[3] H. Kawai, D.C. Lewellen and H. Tye, Phys. Rev. Lett. **57** (1986) 1832, Erratum-*ibid.* **58** (1987) 429; Nucl. Phys. **B288** (1987) 1; W. Lerche, D. Lust and A.N. Schellekens, Nucl. Phys. **B287** (1987) 477; I. Antoniadis, C.P. Bachas and C. Kounnas, Nucl. Phys. **B289** (1987) 87.

[4] J. Maldacena, Adv. Theor. Math. Phys. **2** (1998) 231, hep-th/9711200.

[5] E. D'Hoker and D. Z. Freedman, *'Supersymmetric gauge theories and the AdS/CFT correspondence'*, Lectures given at TASI in Elementary Particle Physics, 'Boulder 2001, Strings, Branes and Extra Dimensions', 3-158, hep-th/0201253.

[6] H. Nishino and S. Rajpoot, *"Vector Multiplet in Non-Adjoint Representation of SO(N)"*, CSULB-PA-07-02 (Feb. 2007); arXiv:0704.2905, *to appear in Phys. Rev. D.*

[7] J. Scherk and J.H. Schwarz, Nucl. Phys. **B153** (1979) 61.

[8] H. Nishino and S. Rajpoot, hep-th/0407203, Phys. Lett. **604B** (2004) 123.